# Verifying the Unification Algorithm in LCF

Lawrence C. Paulson
Computer Laboratory
Corn Exchange Street
Cambridge CB2 3QG
England

**Abstract.** Manna and Waldinger's theory of substitutions and unification has been verified using the Cambridge LCF theorem prover. A proof of the monotonicity of substitution is presented in detail, as an example of interaction with LCF. Translating the theory into LCF's domain-theoretic logic is largely straightforward. Well-founded induction on a complex ordering is translated into nested structural inductions. Correctness of unification is expressed using predicates for such properties as idempotence and most-generality. The verification is presented as a series of lemmas. The LCF proofs are compared with the original ones, and with other approaches. It appears difficult to find a logic that is both simple and flexible, especially for proving termination.

# Contents

# 1 Introduction

Manna and Waldinger have derived a unification algorithm by proving that its specification can be satisfied, to illustrate their technique for program synthesis [17]. They present the proof in detail so that it can be mechanized. The proof, which also constitutes a verification of the unification algorithm, relies on a substantial theory of substitutions, consisting of twenty-three propositions and corollaries. Using the interactive theorem prover LCF [12], I have verified both the unification algorithm and the theory of substitutions.

The project has grown too large to describe in a single paper. This paper is a survey, discussing the main aspects and mentioning papers where you can find more details. The proof is not entirely beautiful. A surprisingly diverse series of problems appeared; some were clumsily solved. I hope to honestly report the difficulties of mechanizing mathematics.

There are few other accounts of large, machine-assisted proofs in the literature. One is the monumental verification of an entire mathematics textbook in the AUTOMATH system [16]. Boyer and Moore have proved a number of difficult theorems using their theorem prover [2, 3].

Although this paper may be read independently, you are advised to read Manna and Waldinger (henceforth MW). I occasionally refer to particular sections of their paper, for example (MW §5). Beware of differences in notation, variable names, and data structures.

The remaining sections present

2. an overview of unification and related concepts;

3. the principles of the LCF theorem prover;

4. differences between MW's informal proof and the LCF one;

5. sample definitions of data types and functions;

6. a detailed proof: the monotonicity of substitution;

7. formalizing the unification algorithm and its statement of correctness;

8. concluding remarks about good and bad aspects of the proof, and prospects for the future.

# 2 Overview of unification

Consider expressions consisting of variables such as $x$ or $y$, constants such as $A$ or $B$, and function applications such as $F[A]$ or $G[A; x]$. Regard a variable $x$ as an empty slot that may be filled with any expression, so long as every occurrence of $x$ gets the same expression. Then two expressions are *unifiable* if they become the same after replacing some of their variables by expressions.

For example, the two expressions $G[A; x]$ and $G[y; F[y]]$ are unified by the substitution $\{x \rightarrow F[A];\ y \rightarrow A\}$, since both become $G[A; F[A]]$. This substitution is called a *unifier*, and can be shown to be the most general possible. A similar pair of expressions, $G[x; x]$ and $G[y; F[y]]$, cannot be unified. For any two expressions, the unification algorithm either produces their most general unifier, or reports that no unifier exists. Unification plays a central role in theorem-proving, polymorphic type-checking [19], the language Prolog [8], and other areas of artificial intelligence [7].

*Note*: If $t$ is an expression such that $t = F[t]$, then putting $t$ for both $x$ and $y$ unifies $G[x; x]$ and $G[y; F[y]]$. This $t$ can be written as the infinite expression $F[F[F[\cdots]]]$, which can easily be formalized in LCF. However, allowing infinite expressions would be a drastic departure from MW's theory.

The underlying theory of substitutions involves a surprising number of functions, relations, and other notions. In the following list, the LCF name of a function is given in UPPER CASE; the theory includes

- the *variables* function (VARS_OF), which gives the set of variables contained in an expression;

- the *occurs-in* relation (OCCS), which determines whether one expression occurs within another;

- the *domain* function, which gives the set of variables affected by a substitution;

- the *range* function, which gives the set of variables that a substitution may introduce;

- the function to *apply* a substitution to an expression (SUBST);

- the *composition* of substitutions (THENS);

- *instances* of substitutions;

- *most-general* and *idempotent* substitutions;

- the function to *unify* two expressions (UNIFY).

The diagram illustrates the relationships among the concepts. The hierarchy in the LCF proof is considerably more complex: dozens of small theories with many interdependencies. Even finite sets, at the bottom of the diagram, involve several LCF theories to specify the set operations and prove the relationships among them.

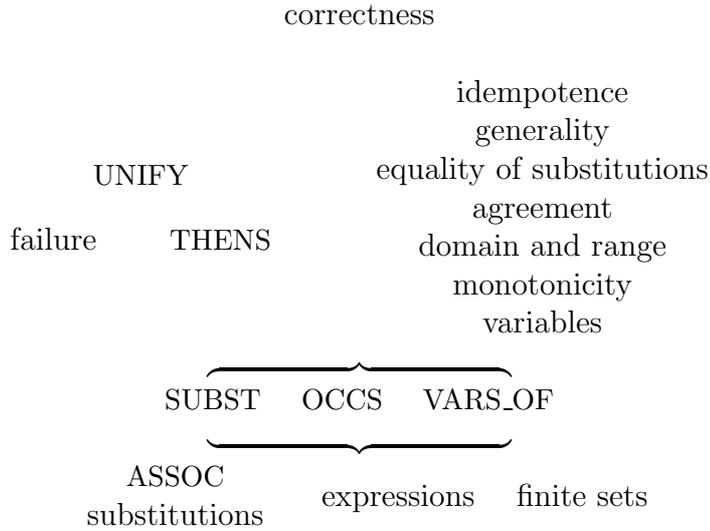

## 3 Overview of LCF

LCF is an interactive, programmable theorem prover for Scott's Logic of Computable Functions. There are several versions of LCF. Cohn has used Edinburgh LCF to prove the equivalence of two semantic definitions of a simple programming language [10]. Mulmuley has used it to automate existence proofs for inclusive predicates, a highly technical aspect of compiler verification [20]. Gordon has extended Cambridge LCF for reasoning about hardware, and proved the correctness of a small computer [14]. This section introduces the principles; tutorials have appeared elsewhere [12, 13]. The unification proof uses Cambridge LCF.

### 3.1 The logic PPLAMBDA

Theorems are proved in the logic PPLAMBDA, which provides the usual predicate calculus formulas, such as conjunctions $P \wedge Q$, disjunctions $P \vee Q$, existentials $\exists x.P$, and logical equivalences $P \iff Q$ [22]. Theorems are proved via inference rules for introducing and eliminating connectives. In this *natural deduction* style, a theorem may depend on assumptions: writing $[P; Q] \vdash R$ means that $R$ is a theorem under the assumptions $P$ and $Q$. A theorem with no assumptions is written $\vdash R$.

A term $t$ may be a constant $C$, a variable $x$, an abstraction $\lambda x.t$, or a combination $t_1\ t_2$. Every term has a *type*, and $t : \alpha$ means that the term $t$ has type $\alpha$. In the semantics of PPLAMBDA, each type denotes a domain (complete partial ordering) instead of a set. Every type includes the "undefined" element $\bot$, which stands for the result of a nonterminating computation.

If $\alpha$ and $\beta$ are types, then $\alpha \to \beta$ is the type of *continuous functions* from $\alpha$ to $\beta$. A function with argument types $\alpha_1$ and $\alpha_2$ and result type $\beta$ is often given the *curried* type $\alpha_1 \to (\alpha_2 \to \beta)$, abbreviated as $\alpha_1 \to \alpha_2 \to \beta$.

The type $\alpha \times \beta$ is the *Cartesian product* of types $\alpha$ and $\beta$; for every $t : \alpha$ and $u : \beta$, the pair $(t, u)$ belongs to $\alpha \times \beta$.

The type *tr* contains the truth values TT, FF, and $\bot$, where TT means true, FF means false, and $\bot$ means undefined. My formalization defines functions AND, OR, and NOT for truth values. These are distinct from the logical connectives $\wedge$, $\vee$, and $\neg$. The infix function = denotes a computable equality test, which is distinct from logical equality, $\equiv$.

The type *tr* represents the kind of truth values that programs often manipulate; logical truth represents provable statements about programs. There are opposing views on how to eliminate this annoying two-tiered notion of truth. Gordon's Higher Order Logic [15] treats arbitrary propositions as truth values. Boyer and Moore [2] allow only computable expressions as formulas, a constructive approach that goes far beyond the demands of intuitionists [18].

The conditional $t \Rightarrow t_1 \mid t_2$, where $t$ has type *tr*, satisfies

$$\begin{aligned} \text{TT} \Rightarrow t_1 \mid t_2 &\equiv t_1 \\ \text{FF} \Rightarrow t_1 \mid t_2 &\equiv t_2 \\ \bot \Rightarrow t_1 \mid t_2 &\equiv \bot \end{aligned}$$

PPLAMBDA allows reasoning about denotational semantics, higher-order functions, infinite data structures, and partial functions. Functional programs can be stated as equations involving lambda expressions. Reasoning about total (always terminating) functions is difficult. Functions must be proved total, and computations proved to terminate, where traditional logics make this implicit. This wastes time and effort of both user and computer. The ugliest reasoning involves *flatness*. A flat type, roughly speaking, is one with no partially defined elements. Examples are the types of natural numbers and finite lists, but not functions or unbounded streams. It is essential to prove flatness in order to use certain functions, such as equality.

PPLAMBDA includes axioms for the fixed-point theory of computation, but not for common data structures like lists. LCF allows you to extend the logical framework, building hierarchies of theories. If you create a theory *nat* of the natural numbers, then you and other users can build new theories on top of *nat*.

## 3.2 The meta-language ML

PPLAMBDA is embedded in LCF's meta-language, ML, which is a functional programming language related to Landin's ISWIM [4]. ML data types include the usual *int* and *bool*, and also *term*, *form*, and *thm*, whose values are PPLAMBDA terms, formulas, and theorems. An axiom is a constant of type *thm*; an inference rule is a function mapping theorems to theorems. Type-checking ensures that a theorem can only be obtained by applying inference rules to axioms.

ML provides simple data structures. A *list* of elements (of the same type) is written $[x_1; \ldots; x_n]$; a *tuple* of elements (of possibly differing types) is written $x, y, \ldots, z$. We shall mostly see lists of theorems or assumptions. If $P, Q_1, \ldots, Q_n$ are formulas, then the *goal* of proving $P$ under the assumptions $Q_i$ is represented as the pair $([Q_1; \ldots; Q_n], P)$. The type *thm* can be viewed as an abstract type: its elements are represented by goals, but access to the representation is restricted.

ML provides a simple form of exception handling. Any expression can signal *failure*, which propagates through enclosing expressions and function calls until it is *trapped*, whereupon an alternative expression is evaluated. For instance, the inference rule for Modus Ponens is the function MP, of type $thm \to thm \to thm$. It fails if the first argument is not an implication, or if the second argument is not the antecedent of the first:

$$\frac{P \implies Q \quad P}{Q}$$

LCF is *programmable*: all commands can be invoked as ML functions. LCF itself contains over five thousand lines of ML, implementing rewriting functions, subgoaling functions (tactics and tacticals), commands for reasoning about recursive data structures, etc. By writing more ML you extend and tailor LCF to the task at hand, perhaps producing a system as large as Mulmuley's [20].

## 3.3 Goal-directed proof

Most LCF proofs are conducted backwards, reducing goals to simpler subgoals. A *tactic* is a function which, given a goal $g$, returns a list of subgoals $[g_1; \ldots; g_n]$, along with a function for proving $g$ as a theorem once the subgoals have been proved.

For example, the inference rule for conjunction introduction,

$$\frac{P \quad Q}{P \wedge Q} \quad ,$$

is provided by the ML function CONJ, which has type $thm \to thm \to thm$. The corresponding tactic is CONJ_TAC. Given a goal $([R_1; \ldots; R_n], P \wedge Q)$, the tactic returns the list of goals $[([R_1; \ldots; R_n], P); ([R_1; \ldots; R_n], Q)]$, along with a function that calls CONJ. Normally we keep the assumptions implicit and simply say that CONJ_TAC reduces $P \wedge Q$ to the goals $P$ and $Q$. The tactic fails if the goal is not a conjunction.

The "discharge rule" for implication introduction,

$$\frac{[P] \, Q}{P \implies Q} \quad ,$$

is provided by the ML function DISCH. The corresponding tactic is DISCH_TAC, which reduces $P \implies Q$ to the goal of proving $Q$ under the assumption $P$ plus any previous assumptions.

Functions, and hence tactics, are first-class values in ML. Tactics can be combined into more powerful ones using operators called *tacticals*, such as THEN, ORELSE, and REPEAT. If $tac_1$ and $tac_2$ are tactics, then $tac_1$ THEN $tac_2$ is a tactic that applies $tac_1$ to its goal, and applies $tac_2$ to the resulting subgoals. The tactic $tac_1$ ORELSE $tac_2$ returns the subgoals given by $tac_1$, applying $tac_2$ if $tac_1$ fails. The tactic REPEAT $tac$ applies $tac$ repeatedly to the goal and its subgoals.

Gordon's tutorial on ML [13] describes how to implement inference rules and tactics for a simple logic. For reading this paper, it is all right if you equate the tacticals THEN, ORELSE, and REPEAT with the notions of sequencing, alternation, and repetition. Cambridge LCF contains additional tacticals for iteration and handling assumptions, and tactics for each inference rule and other reasoning primitives [23]. Nearly any tactic can be expressed in terms of other tactics and tacticals, without requiring low-level ML code that explicitly builds lists of subgoals. Cambridge LCF uses same approach for rewriting: simplifiers are expressed in terms of rewriting primitives and operators for combining them [21].

## 3.4 Recursive data structures

PPLAMBDA can express theories of a variety of common data structures, such as the natural numbers, lists and trees. Data structures can be infinite (lazy) or finite, and mutually recursive; sometimes you can impose equational constraints and produce a quotient type [24]. The rule of structural induction is derived from PPLAMBDA's rule of fixed-point induction.

Many recursive data types can be described as a set of constructor functions, each taking a number of arguments of specified types. This determines the constants, types, and axioms needed to define the type in a PPLAMBDA theory. The construction of the theory and the derivation of induction can be invoked through Cambridge LCF commands; examples appear below. These commands are descended from the structural induction package written by Milner for Edinburgh LCF [9]. In a similar spirit, the Boyer-Moore theorem prover provides a command for defining a new data structure by introducing a number of functions [2].

# 4 Differences between the formal and informal theories

A key question about any mechanical theorem prover is: how comfortably does it accomodate a mathematician's informal reasoning? LCF can handle the proofs of Manna and Waldinger (MW), but not as originally stated. The LCF proofs differ in the underlying logic and in the data structure for expressions. In my first attempt to formalize their proof, I also used a different data type for sets and substitutions.

MW's theory fits together tightly. On most occasions when I made even a slight change in the statement of a theorem, this caused surprising problems later. As there were false attempts using mutually recursive expressions, and using lists instead of sets, large parts of the theory were verified two or three times. Some of the early proofs were repeated many times, and now are nicely polished. It is interesting that both of the major false starts involved the choice of data structures. The final proof closely resembles MW's, except in the data structure for expressions and the continual fiddling with $\bot$.

MW's aim is *deductive synthesis* – deriving a program from a proof that its specification can be satisfied. My proof does not synthesize a program; I state the unification algorithm and then prove it. Conducting MW's proof in a logic of constructive type theory would synthesize a correct program [18].

## 4.1 Logical framework

MW's proof is presented in an ordinary first-order logic, with several fundamental differences from PPLAMBDA:

- variables range over sets, not domains;

- all functions are total, while PPLAMBDA allows partial functions and functionals;

- there are no types, while PPLAMBDA has a polymorphic type system;

- the induction principle is well-founded induction, while PPLAMBDA uses structural and fixed-point induction.

PPLAMBDA's explicit handling of termination is a nuisance for functions that obviously terminate. The LCF formalization is cluttered with theorems about the termination of simple functions, not found in MW. On the other hand, the termination of the unification algorithm is a deep theorem that depends on the partial correctness of the algorithm. PPLAMBDA accepts unification as a partial function, so that the necessary lemmas can be proved.

## 4.2 Data structure for expressions

MW define the unification algorithm for so-called *l*-expressions. An expression is either a constant, a variable, or a function application $f \cdot [l_1; \ldots; l_n]$, containing a list of expressions. An *l*-expression is either an expression or a list of *l*-expressions. Lists are built up, in Lisp style, from the empty list $[]$ using the "cons" operator $\circ$. Thus $[l_1; \ldots; l_n]$ abbreviates $l_1 \circ (\cdots \circ (l_n \circ []))$.

MW mention both "*l*-expressions" and "lists of *l*-expressions," which suggests that there are two *mutually recursive* types, not a single type. In the first attempt, I spent several months deriving the mutual induction rule and implementing an LCF program for defining mutually recursive data structures. After proving a few theorems in the framework, I realized that a simpler data structure would shorten the proofs without sacrificing generality.

There is no need to provide both constant and function symbols. Nor do we need lists of expressions, which MW use as argument lists of functions. In the lambda-calculus, a function can have only one argument; the effect of multiple arguments is achieved by *currying*, where a function returns a function as its result. Instead of writing $G[A; x]$, we can write $(G(A))(x)$.

These considerations lead us to the data structure used in the LCF proofs. A *term* is either

**a** *constant* such as $A$, $B$, $F$, or $G$, or

**a** *variable* such as $x$ or $y$, or

**a** *combination* such as $(t_1 t_2)$, where $t_1$ and $t_2$ are terms.

I call these structures "terms" because they are similar to PPLAMBDA terms, and continue to use the word "expression" when discussing MW's theory. Let us hope this will not cause confusion. If you would like to see a more familiar representation of expressions, take these terms to denote binary trees or Lisp S-expressions.

## 4.3 Sets and substitutions

To show that a function $f$ always terminates, it suffices to show that for any call $f(x)$, the argument $y$ in each recursive call $f(y)$ is smaller than $x$ in some well-founded ordering [2]. Most commonly, $y$ is a substructure of $x$. Expressed as a recursive function on terms, unification involves a call where the arguments may become bigger than the original terms. When this happens, the set of variables in the terms can be shown to decrease. The proof that unification terminates relies on a well-founded ordering involving both terms and sets. MW reason extensively about finite sets of variables.

It seems clear that any statement about finite sets can be translated in terms of lists, using Lisp-like functions to implement set operations. Thus APPEND represents union, MEMBER represents the membership predicate, etc. This representation complicates proofs. The union of sets satisfies several algebraic laws:

$$
\begin{aligned}
a \cup b &= b \cup a & \text{commutative} \\
a \cup a &= a & \text{idempotent} \\
(a \cup b) \cup c &= a \cup (b \cup c) & \text{associative} \\
a \cup (b \cap c) &= (a \cup b) \cap (a \cup c) & \text{distributive}
\end{aligned}
$$

Of these, APPEND enjoys only the associative law. The other list operations fare little better.

I reformulated MW's theory using lists and verified most of it. Reasoning about lists was so awkward that I did not attempt the final section of the proof (§21), where MW introduce the well-founded ordering. So I had to revert to sets. It was not obvious how to define sets in PPLAMBDA, particularly as they would have to allow proof by structural induction. Once I worked this out, it was straightforward to define the familiar set operations, and prove their properties. A fundamental one is *extensionality*:

$$(a = b) \iff (\forall x.\, x \in a \iff x \in b).$$

Finite sets are a *quotient type* — they can be defined as equivalence classes of finite lists, where the order and multiplicity of elements is ignored. The equivalence relation can be stated as two equations, taking care to avoid inconsistencies concerning the derivation of structural induction from fixed-point induction [24].

If successful, unification produces a substitution, such as $\{x \to F[A];\ y \to A\}$. The obvious representation of a substitution is a list of pairs: $[(x, F[A]);\ (y, A)]$. However, MW impose additional requirements on substitutions. They *forbid* trivial ones like $\{x \to x\}$, and ambigous ones like $\{x \to A; x \to B\}$. Furthermore, MW state that two substitutions are equal if they agree on all expressions. They rely heavily on this definition of equality, making it impractical to carry out their proofs using lists of pairs.

This type seems harder to define than sets, as it requires error values. My formalization allows all substitutions. It identifies $\{x \to x\}$ with the empty substitution $\{\}$. It resolves ambiguous substitutions by identifying $\{x \to A; x \to B\}$ with $\{x \to A\}$. Regrettably, the order of elements matters in this case; substitutions are not sets of pairs. However, this formalization simplifies MW's theory; their subtraction function (MW §8) is not needed.

## 4.4 The induction principle

MW's logic provides for induction on any well-founded ordering. I know of no LCF derivation of this extremely general principle. Boyer and Moore use well-founded induction, but in a restricted form that can be reduced to one or more appeals to induction on the natural numbers [2]. Most of MW's inductive arguments are ordinary structural induction. Their one difficult well-founded induction can be reformulated as two structural inductions: one on the natural numbers, one on terms.

Well-founded induction provides a clear and concise notation for an inductive argument. Using their ordering $\prec_{\text{un}}$, MW prove several theorems of the form $(t, t') \prec_{\text{un}} (u, u')$. In my simulation of well-founded induction on $\prec_{\text{un}}$, these become large formulas containing multiple occurrences of $t$, $t'$, $u$, and $u'$.

# 5 Constructing theories in LCF

Until now we have been discussing LCF proofs in a general sense. Now let us examine the commands used to define some data types and functions in the formalization. One LCF theory defines terms; another defines substitutions. Function definitions reside on various theories. The theory hierarchy is a directed acyclic graph, reflecting the dependencies and other relationships among the functions and types. If theory $A$ depends on theory $B$, then $B$ is called a *parent* of $A$, and must be declared to LCF when $A$ is created.

## 5.1 Expressions

The data structure for expressions, the PPLAMBDA type *term*, consists of *constants*, *variables*, and *combinations* of terms. The formalization requires types *const* and *var* for constant and variable symbols. We can start a new LCF theory and store these types by issuing the commands

    *new_theory* 'term';;
    *new_type* 0 'const';;
    *new_type* 0 'var';;
    *new_type* 0 'term';;

Nothing more need be said about *const* and *var*. A single command introduces a recursive structure on *term*:

  *struct_axm* ("*: term*", 'strict',
   ['CONST', ["*c* : *const*"];
   'VAR', ["*v* : *var*"];
   'COMB', ["*t*$_1$ : *term*"; "*t*$_2$ : *term*"]] );;

This causes LCF to declare the constructor functions with appropriate types:

$$\begin{aligned}\text{CONST} &: \mathit{const} \to \mathit{term} \\ \text{VAR} &: \mathit{var} \to \mathit{term} \\ \text{COMB} &: \mathit{term} \to \mathit{term} \to \mathit{term}\end{aligned}$$

It asserts axioms stating that these are distinct, one-to-one, and so forth [24]. To exclude infinite terms, the constructor functions are all *strict*: for instance, CONST $\bot \equiv \bot$.

The LCF tactic STRUCT_TAC derives structural induction for any type defined using *struct_axm*. On a goal $\forall t.\, P(t)$, induction on the term $t$ produces four subgoals: $t$ may be undefined; $t$ may be a constant or variable; $t$ may be the combination of $t_1$ and $t_2$, which satisfy induction hypotheses. The subgoals include assumptions that the substructures are defined. The induction tactic for terms corresponds to the rule

$$\frac{\begin{array}{rl} & P(\bot) \\ [c \not\equiv \bot] & P(\text{CONST}\, c) \\ [v \not\equiv \bot] & P(\text{VAR}\, v) \\ [P(t_1);\ P(t_2);\ t_1 \not\equiv \bot;\ t_2 \not\equiv \bot] & P(\text{COMB}\, t_1 t_2) \end{array}}{\forall t.\, P(t)}$$

## 5.2   A recursive function on expressions

MW define the *occurs-in* relation on two expressions, where $t$ occurs in $u$ if $t$ is a subexpression of $u$. In my formalization, the reflexive form of this relation is called OCCS_EQ and the anti-reflexive form is called OCCS. These can be defined as mutually recursive functions taking arguments of type *term* and returning a truth-valued result. They are declared in a new theory called OCCS. By making them *infix* functions, we can write $t$ OCCS $u$ instead of OCCS $t\, u$:

  *new_theory* 'OCCS';;
  *new_curried_infix* ('OCCSEQ', ": *term* $\to$ *term* $\to$ *tr*");;
  *new_curried_infix* ('OCCS', ":*term* $\to$ *term* $\to$ *tr*");;

We can define OCCS and OCCS_EQ via two axioms. The command *new_axiom* stores a named formula as an axiom of the current theory.

  *new_axiom* ('OCCSEQ',
   " $\forall t\, u.\ t$ OCCS_EQ $u \equiv (t = u)$ OR $(t$ OCCS $u)$");;

```
    new_axiom ('OCCS_CLAUSES',
      "∀t.  t OCCS ⊥ ≡ ⊥ ∧
        (∀c. c ≢ ⊥ ⟹  t OCCS (CONST c) ≡ FF) ∧
        (∀v. v ≢ ⊥ ⟹  t OCCS (VAR v) ≡ FF) ∧
        (∀t₁t₂. t₁ ≢ ⊥ ⟹ t₂ ≢ ⊥ ⟹
           t OCCS (COMB t₁t₂) ≡ (t OCCS_EQ t₁) OR (t OCCS_EQ t₂))");;
```

The definition of OCCS consists of four equations, which exhaust all possible inputs. This is reminiscent of programming in Prolog [8] or HOPE [6], and has become standard LCF style [9]. Burstall recommended it long ago, in his still timely discourse on how to formulate and prove theorems by structural induction [5]. Such function definitions are easy to read because they present actual patterns of computation. They are ready for use as rewrite rules, providing symbolic execution. The old-fashioned approach requires conditional expressions involving discriminator and destructor functions.

The numerous *definedness assertions*, such as $c \not\equiv \bot$, prevent the left sides of the equations from overlapping. If omitted, $c \equiv \bot$ would give the contradiction

$$\bot \;\equiv\; t \text{ OCCS } \bot \;\equiv\; t \text{ OCCS } (\text{CONST } c) \;\equiv\; \text{FF} .$$

These essential assertions can be hidden with the help of special quantifers, but still must be manipulated in proofs.

In order to allow structural induction on expressions, MW take as an axiom that the relation OCCS is well-founded, implying that it is anti-reflexive and anti-symmetric. Structural induction is derived differently in LCF, where OCCS is just another function. Either way, OCCS is a partial ordering. In LCF it can be proved to be anti-reflexive, anti-symmetric, and transitive, by induction on terms. In fact, OCCS is related to the less-than relation on the natural numbers, whose partial ordering properties have essentially the same LCF proofs.

## 5.3  Substitutions

For MW, a substitution is an abstract type, represented as a set of pairs. Given a substitution, the *apply function* transforms one expression into another, using the substitution to replace variables by expressions (MW §5). The function traverses both expressions (to build new ones) and substitutions (to find variables). My formalization splits these roles into separate functions ASSOC and SUBST.

Substitutions can be defined independently from terms. I use a postfix type operator *alist*, taking two type arguments. If $\alpha$ and $\beta$ are types, then $(\alpha, \beta)alist$ is a type of association lists of the form

$$\{key_1 \to val_1; \cdots; key_n \to val_n\},$$

where $key_i : \alpha$ and $val_i : \beta$ for $i = 1, \ldots n$. The constructors are

$$\begin{aligned}\text{ANIL} &: (\alpha, \beta)alist \\ \text{ACONS} &: \alpha \to \beta \to (\alpha, \beta)alist \to (\alpha, \beta)alist\end{aligned}$$

where ANIL stands for the empty substitution {}, while ACONS extends an existing substitution. If $key : \alpha$ and $val : \beta$, then

$$\text{ACONS } key\ val\ \{key_1 \to val_1; \cdots; key_n \to val_n\}$$
$$\equiv \{key \to val; key_1 \to val_1; \cdots; key_n \to val_n\} \ .$$

The function call ASSOC *nilval key s* searches the substitution *s* for the given *key*, and returns the associated value. If the key is not present, ASSOC returns the argument *nilval*. Its theory depends on parent theories of alists, equality, and truth values:

> *new_theory* 'ASSOC';;
> *new_parent* 'alist';;
> *new_parent* 'eq';;
> *new_parent* 'tr';;
> *new_constant* ('ASSOC', ": $\beta \to \alpha \to (\alpha, \beta)alist \to \beta$");;

Its definition is again a conjunction of clauses, one for each kind of input:

> *new_axiom* ('ASSOC_CLAUSES',
> "$\forall nilval : \beta.\ \forall key' : \alpha.$
>   ASSOC *nilval key'* $\bot \equiv \bot \land$
>   ASSOC *nilval key'* ANIL $\equiv nilval \land$
>   $(\forall key\ val\ s.\ key \not\equiv \bot \implies val \not\equiv \bot \implies s \not\equiv \bot \implies$
>     ASSOC *nilval key'* (ACONS *key val s*) $\equiv$
>     $(key' = key) \Rightarrow val\ |\ $ASSOC *nilval key' s$)$");;

SUBST is an infix function; the invocation *t* SUBST *s* applies the substitution *s* to the term *t*. If *t* is a constant, it is returned; if *t* is a variable, it is looked up in *s*; if *t* is a combination, then SUBST calls itself on the two subterms. SUBST is defined using the functional TERM_REC for structural recursion on terms. Such functionals appear throughout the formalization, but could be removed easily. I wish to avoid digressing into this area, which Burge [4] describes with many examples:

*new_theory* 'SUBST';;
*new_parent* 'TERM_REC';;
*new_parent* 'ASSOC';;
*new_curried_infix* ('SUBST', ": $term \to (var, term)alist \to term$ ");;
*new_axiom* ('SUBST',
  "$\forall ts.\ t$ SUBST $s \equiv$ TERM_REC(CONST, $(\lambda v.$ ASSOC(VAR $v$)$v\ s$), COMB) $t$ ");;

Since TERM_REC can be seen as a generalization of the APL reduce operator, it is not surprising that SUBST has a one-line definition. For most reasoning it is preferable to have a definition of SUBST in the form used for OCCS and ASSOC, namely a conjunction of equations. Here the equations are not taken as an axiom, but proved as a theorem, using rewriting to unfold TERM_REC. The *prove_thm* command applies a tactic to a formula, storing the resulting theorem if successful. The tactic is just a call to REWRITE_TAC, a standard rewriting tactic. Sorry for omitting the details; we shall see more substantial proofs shortly.

>        *prove_thm* (‘SUBST_CLAUSES’,
>            “∀s. ⊥ SUBST s ≡ ⊥ ∧
>                (∀c. c ≢ ⊥ ⟹ (CONST c) SUBST s ≡ CONST c) ∧
>                (∀v. v ≢ ⊥ ⟹ (VAR v) SUBST s ≡ ASSOC(VAR v)v s) ∧
>                (∀t₁t₂. t₁ ≢ ⊥ ⟹ t₂ ≢ ⊥ ⟹
>                    (COMB t₁t₂) SUBST s ≡ COMB(t₁ SUBST s)(t₂ SUBST s))”,
>        REWRITE_TAC [SUBST; TERM_REC_CLAUSES]);;

# 6  The monotonicity of substitution

An elementary theorem of Manna and Waldinger (MW §5) asserts that substitution is monotonic relative to the occurs-in relation. If the expression $t$ occurs in $u$, then applying any substitution $s$ preserves this relationship, so $t$ SUBST $s$ occurs in $u$ SUBST $s$. This section presents the machine proof using smaller steps than necessary, to illustrate the internal and external workings of LCF. The first step is to invoke LCF and create a new theory to hold theorems about monotonicity. Its parents are the theories of functions SUBST and OCCS:

> *new_theory* ‘mono’;;
> *new_parent* ‘SUBST’;;
> *new_parent* ‘OCCS’;;

Now that LCF knows the types of OCCS and SUBST, we can state the monotonicity of substitution as the goal. The interactive subgoal package keeps track of the current goal and others, both solved and unsolved.

> *set_goal*([],
>     “∀s. s ≢ ⊥ ⟹ ∀t. t ≢ ⊥ ⟹
>         ∀u. t OCCS u ≡ TT ⟹ (t SUBST s) OCCS (u SUBST s) ≡ TT”);;

A little thought reveals that structural induction on the variable $u$ will open up the recursive definitions of OCCS and SUBST. If we type

> *expand* (STRUCT_TAC ‘term’ [] “u”);;

then LCF prints four subgoals, with assumptions underneath in square brackets:

$t$ OCCS (COMB $t_1t_2$) ≡ TT ⟹ ($t$ SUBST $s$) OCCS ((COMB $t_1t_2$) SUBST $s$) ≡ TT
  [$s ≢ ⊥$; $t ≢ ⊥$; $t_1 ≢ ⊥$; $t_2 ≢ ⊥$;
   $t$ OCCS $t_1$ ≡ TT ⟹ ($t$ SUBST $s$) OCCS ($t_1$ SUBST $s$) ≡ TT;
   $t$ OCCS $t_2$ ≡ TT ⟹ ($t$ SUBST $s$) OCCS ($t_2$ SUBST $s$) ≡ TT]

$t$ OCCS (VAR $v$) ≡ TT ⟹ ($t$ SUBST $s$) OCCS ((VAR $v$) SUBST $s$) ≡ TT
  [$s ≢ ⊥$; $t ≢ ⊥$; $v ≢ ⊥$]

$t$ OCCS (CONST $c$) ≡ TT ⟹ ($t$ SUBST $s$) OCCS ((CONST $c$) SUBST $s$) ≡ TT
  [$s ≢ ⊥$; $t ≢ ⊥$; $c ≢ ⊥$]

$t$ OCCS ⊥ ≡ TT ⟹ ($t$ SUBST $s$) OCCS (⊥ SUBST $s$) ≡ TT
  [$s ≢ ⊥$; $t ≢ ⊥$]

As usual, induction has given goals that can be simplified by unfolding function definitions. The LCF tactic ASM_REWRITE_TAC uses a list of theorems to rewrite the goal [21]. The axiom OCCS_CLAUSES is a conjunction containing several rewrite rules, one for $\bot$ and one for each term constructor:

$$\vdash t \text{ OCCS } \bot \equiv \bot$$
$$\vdash c \not\equiv \bot \implies t \text{ OCCS } (\text{CONST } c) \equiv \text{FF}$$
$$\vdash v \not\equiv \bot \implies t \text{ OCCS } (\text{VAR } v) \equiv \text{FF}$$
$$\vdash t_1 \not\equiv \bot \implies t_2 \not\equiv \bot \implies$$
$$t \text{ OCCS } (\text{COMB } t_1 t_2) \equiv (t \text{ OCCS\_EQ } t_1) \text{ OR } (t \text{ OCCS\_EQ } t_2)$$

(A theorem of the form $A_1 \implies \cdots \implies A_n \implies t \equiv u$ is an *implicative rewrite*; the rewriting tactic replaces $t$ by $u$ whenever it can prove every $A_i$.)

The $\bot$ subgoal is the first one to tackle. Rewriting can transform its antecedent, $t \text{ OCCS } \bot \equiv \text{TT}$, to the contradiction $\bot \equiv \text{TT}$, solving the goal. We need only type

*expand* (ASM_REWRITE_TAC [OCCS_CLAUSES]);;

whereupon LCF prints the remaining three subgoals. The CONST and VAR ones are solved similarly. The only remaining subgoal, for terms of the form COMB, has no trivial proof by contradiction. We will have to unfold the function SUBST; the goal also involves OCCS_EQ, via the COMB clause for OCCS. Expanding with the tactic

ASM_REWRITE_TAC [OCCS_CLAUSES; OCCSEQ; SUBST_CLAUSES]

produces

$$((t = t_1) \text{ OR } (t \text{ OCCS } t_1)) \text{ OR } ((t = t_2) \text{ OR } (t \text{ OCCS } t_2)) \equiv \text{TT} \implies$$
$$(t \text{ SUBST } s) \text{ OCCS } (\text{COMB}(t_1 \text{ SUBST } s)(t_2 \text{ SUBST } s)) \equiv \text{TT}$$
$$[s \not\equiv \bot;\ t \not\equiv \bot;\ t_1 \not\equiv \bot;\ t_2 \not\equiv \bot;$$
$$t \text{ OCCS } t_1 \equiv \text{TT} \implies (t \text{ SUBST } s) \text{ OCCS } (t_1 \text{ SUBST } s) \equiv \text{TT}$$
$$t \text{ OCCS } t_2 \equiv \text{TT} \implies (t \text{ SUBST } s) \text{ OCCS } (t_2 \text{ SUBST } s) \equiv \text{TT}]$$

Why doesn't the tactic unfold the ... OCCS (COMB ...) in the consequent? It can not prove the antecedents of the COMB rule for OCCS without a theorem stating that SUBST is total. Forgetting to include enough rewrite rules is a common error. It is particularly puzzling when a totality theorem has been left out. Nor should you include all the rewrite rules you have, for this can slow down the rewriting tactic and blow up the goal to enormous size. Here I deliberately omitted SUBST_TOTAL to prevent the unfolding of the consequent. For the time being, let's worry just about the antecedent.

It is helpful to replace the function OR by the connective $\vee$ whenever possible, via the theorem OR_EQ_TT_IFF:

$$\vdash p \not\equiv \bot \implies (p \text{ OR } q \equiv \text{TT} \iff p \equiv \text{TT} \vee q \equiv \text{TT})$$

The theorem EQUAL_IFF_EQ allows replacing the function $=$ by the predicate $\equiv$. It holds for any flat type $\alpha$, requiring an ugly formulation that I will not discuss in detail:

$$\vdash \text{FLAT}(\bot : \alpha) \implies \forall x\, y : \alpha.\ x \not\equiv \bot \implies y \not\equiv \bot \implies ((x = y \equiv \text{TT}) \iff x \equiv y)$$

These theorems involve the connective $\iff$; they are *formula* rewrites, for replacing a formula by an equivalent formula. Given these theorems, plus totality theorems to prove their antecedents, ASM_REWRITE_TAC removes every OR and $=$ from the goal:

ASM_REWRITE_TAC [OR_EQ_TT_IFF; EQUAL_IFF_EQ;
    TERM_FLAT_UU; OCCS_TOTAL; EQUAL_TOTAL; OR_TOTAL]

The result is

$$((t \equiv t_1 \implies (t_1\ \text{SUBST}\ s)\ \text{OCCS}\ (\text{COMB}(t_1\ \text{SUBST}\ s)(t_2\ \text{SUBST}\ s)) \equiv \text{TT}) \wedge$$
$$(t\ \text{OCCS}\ t_1 \equiv \text{TT} \implies (t\ \text{SUBST}\ s)\ \text{OCCS}\ (\text{COMB}(t_1\ \text{SUBST}\ s)(t_2\ \text{SUBST}\ s)) \equiv \text{TT})) \wedge$$
$$(t \equiv t_2 \implies (t_2\ \text{SUBST}\ s)\ \text{OCCS}\ (\text{COMB}(t_1\ \text{SUBST}\ s)(t_2\ \text{SUBST}\ s)) \equiv \text{TT}) \wedge$$
$$(t\ \text{OCCS}\ t_2 \equiv \text{TT} \implies (t\ \text{SUBST}\ s)\ \text{OCCS}\ (\text{COMB}(t_1\ \text{SUBST}\ s)(t_2\ \text{SUBST}\ s)) \equiv \text{TT})$$
$$[s \not\equiv \bot;\ t \not\equiv \bot;\ t_1 \not\equiv \bot;\ t_2 \not\equiv \bot;$$
$$t\ \text{OCCS}\ t_1 \equiv \text{TT} \implies (t\ \text{SUBST}\ s)\ \text{OCCS}\ (t_1\ \text{SUBST}\ s) \equiv \text{TT};$$
$$t\ \text{OCCS}\ t_2 \equiv \text{TT} \implies (t\ \text{SUBST}\ s)\ \text{OCCS}\ (t_2\ \text{SUBST}\ s) \equiv \text{TT}]$$

The rewriting tactic has returned a *conjunction*, not a disjunction. It has repeatedly applied the De Morgan law that $(P \vee Q) \implies R$ is equivalent to $(P \implies R) \wedge (Q \implies R)$, causing a case split. In the implication $t \equiv t_1 \implies \cdots$, it has used the antecedent to replace $t$ by $t_1$ in the consequent. Repeatedly applying CONJ_TAC and DISCH_TAC breaks apart the conjunctions and implications:

REPEAT (CONJ_TAC ORELSE DISCH_TAC)

Here are two of the four subgoals. The other two are similar, but have $t_2$ instead of $t_1$ in the last assumption.

$$(t\ \text{SUBST}\ s)\ \text{OCCS}\ (\text{COMB}(t_1\ \text{SUBST}\ s)(t_2\ \text{SUBST}\ s)) \equiv \text{TT}$$
$$[s \not\equiv \bot;\ t \not\equiv \bot;\ t_1 \not\equiv \bot;\ t_2 \not\equiv \bot;$$
$$t\ \text{OCCS}\ t_1 \equiv \text{TT} \implies (t\ \text{SUBST}\ s)\ \text{OCCS}\ (t_1\ \text{SUBST}\ s) \equiv \text{TT};$$
$$t\ \text{OCCS}\ t_2 \equiv \text{TT} \implies (t\ \text{SUBST}\ s)\ \text{OCCS}\ (t_2\ \text{SUBST}\ s) \equiv \text{TT};$$
$$t\ \text{OCCS}\ t_1 \equiv \text{TT}]$$

$$(t_1\ \text{SUBST}\ s)\ \text{OCCS}\ (\text{COMB}(t_1\ \text{SUBST}\ s)(t_2\ \text{SUBST}\ s)) \equiv \text{TT}$$
$$[s \not\equiv \bot;\ t \not\equiv \bot;\ t_1 \not\equiv \bot;\ t_2 \not\equiv \bot;$$
$$t\ \text{OCCS}\ t_1 \equiv \text{TT} \implies (t\ \text{SUBST}\ s)\ \text{OCCS}\ (t_1\ \text{SUBST}\ s) \equiv \text{TT};$$
$$t\ \text{OCCS}\ t_2 \equiv \text{TT} \implies (t\ \text{SUBST}\ s)\ \text{OCCS}\ (t_2\ \text{SUBST}\ s) \equiv \text{TT}$$
$$t \equiv t_1]$$

We can approach the second goal by opening up the $\cdots$ OCCS (COMB $\cdots$) in the consequent, including SUBST_TOTAL this time:

ASM_REWRITE_TAC [OCCS_CLAUSES; OCCSEQ; SUBST_TOTAL]

Replacing OR by $\vee$ and = by $\equiv$ solves the resulting goal; omitting those rules lets us see the intermediate stage:

$(((t_1 \text{ SUBST } s) = (t_1 \text{ SUBST } s)) \text{ OR } ((t_1 \text{ SUBST } s) \text{ OCCS } (t_1 \text{ SUBST } s)))\text{OR}$
$(((t_1 \text{ SUBST } s) = (t_2 \text{ SUBST } s)) \text{ OR } ((t_1 \text{ SUBST } s) \text{ OCCS } (t_2 \text{ SUBST } s))) \equiv \text{TT}$
$[s \not\equiv \bot;\ t \not\equiv \bot;\ t_1 \not\equiv \bot;\ t_2 \not\equiv \bot;$
$\quad t \text{ OCCS } t_1 \equiv \text{TT} \implies (t \text{ SUBST } s) \text{ OCCS } (t_1 \text{ SUBST } s) \equiv \text{TT};$
$\quad t \text{ OCCS } t_2 \equiv \text{TT} \implies (t \text{ SUBST } s) \text{ OCCS } (t_2 \text{ SUBST } s) \equiv \text{TT};$
$\quad t \equiv t_1]$

Try convincing yourself that the other goal is solved similarly. The induction hypothesis, an implicative rewrite, replaces one of operands of OR by TT. Unfolding OR and = finishes the proof. Once all the subgoals have been solved, the subgoal package assembles the bits of the proof, printing the intermediate and final theorems.

...... $\vdash ((t = t_1) \text{ OR } (t \text{ OCCS } t_1)) \text{ OR } ((t = t_2) \text{ OR } (t \text{ OCCS } t_2)) \equiv \text{TT} \implies$
$\quad (t \text{ SUBST } s) \text{ OCCS } (\text{COMB}(t_1 \text{ SUBST } s)(t_2 \text{ SUBST } s)) \equiv \text{TT}$
...... $\vdash ((t = t_1) \text{ OR } (t \text{ OCCS } t_1)) \text{ OR } ((t = t_2) \text{ OR } (t \text{ OCCS } t_2)) \equiv \text{TT} \implies$
$\quad (t \text{ SUBST } s) \text{ OCCS } ((\text{COMB } t_1 t_2) \text{ SUBST } s) \equiv \text{TT}$
...... $\vdash (t \text{ OCCS\_EQ } t_1) \text{ OR } (t \text{ OCCS\_EQ } t_2) \equiv \text{TT} \implies$
$\quad (t \text{ SUBST } s) \text{ OCCS } ((\text{COMB } t_1 t_2) \text{ SUBST } s) \equiv \text{TT}$
...... $\vdash t \text{ OCCS } (\text{COMB } t_1 t_2) \equiv \text{TT} \implies$
$\quad (t \text{ SUBST } s) \text{ OCCS } ((\text{COMB } t_1 t_2) \text{ SUBST } s) \equiv \text{TT}$
$\vdash \forall s.\ s \not\equiv \bot \implies \forall t.\ t \not\equiv \bot \implies$
$\quad \forall u.\ t \text{ OCCS } u \equiv \text{TT} \implies (t \text{ SUBST } s) \text{ OCCS } (u \text{ SUBST } s) \equiv \text{TT}$

I have tried to illustrate how the interactive search for a proof might proceed. After all, the purpose of LCF is not to impose some unwieldy formalism onto our intuitions, but to aid in the discovery of correct proofs. MW do not present the proofs of most of their theorems about substitutions; I rediscovered them with LCF's help.

The proof can be greatly shortened. Supplying all the rewrite rules above, a single call to the tactic ASM_REWRITE_TAC handles all the rewriting. The tactic does not require the goal to be broken up by calling CONJ_TAC or DISCH_TAC beforehand. Using the tactical THEN, the shortened proof can be stated as the composite tactic

```
STRUCT_TAC 'term' [] "u" THEN
ASM_REWRITE_TAC
    [OCCS_CLAUSES; OCCSEQ; SUBST_CLAUSES;
      OR_EQ_TT_IFF; EQUAL_IFF_EQ; TERM_FLAT_UU;
      OCCS_TOTAL; EQUAL_TOTAL; SUBST_TOTAL;
      OR_TOTAL]);;
```

Many LCF proofs consist of nothing but induction followed by rewriting, though few work out this nicely. The proof that OCCS is transitive begins similarly, but some of the equalities are the wrong way round for rewriting, requiring extra steps to reorient them.

# 7 The unification algorithm in LCF

Let us pass by the rest of the theory of substitutions, and concentrate on unification itself. After defining the composition of substitutions, and formalizing a notion of failure, the unification algorithm can be stated as a recursive function. I presume that you have seen enough details of LCF; the remaining examples emphasize logical aspects of the formalization, comparing it with the proof by Manna and Waldinger (MW). Again, beware of differences in notation.

For quantifying over defined values, let $\forall_D x. P$ abbreviate $\forall x. x \not\equiv \bot \implies P$, and let $\exists_D x. P$ abbreviate $\exists x. x \not\equiv \bot \land P$. Implicit universal quantifiers (over free variables in formulas) cover defined values only. This hides irrelevant clutter about $\bot$.

## 7.1 Composition of substitution

Many implementations of unification build up a substitution one variable at a time [7]. MW verify a simpler algorithm involving the operation of composing two substitutions. My symbol for composition is the infix operator THENS (by analogy with the tactical THEN). If $r$ and $s$ are substitutions, then MW (§6) define the composition $r$ THENS $s$ to be the substitution satisfying, for all expressions $t$,

$$t \text{ SUBST } (r \text{ THENS } s) \equiv (t \text{ SUBST } r) \text{ SUBST } s \ .$$

This "definition" does not show that such a substitution exists or is unique. In my formalization, the equation holds for variables by construction, and then is shown to hold for all expressions. I now prefer a simpler definition of THENS, taking as axioms two theorems of the formalization:

$$\text{ANIL THENS } s \equiv s$$
$$(\text{ACONS } v\ t\ r) \text{ THENS } s \equiv \text{ACONS } v(t \text{ SUBST } s)(r \text{ THENS } s)$$

The second equation generalizes the Addition-Composition Proposition (MW §8). You should be careful with definitions by recursion equations; since substitutions are not constructed uniquely from ANIL and ACONS, equations may be inconsistent.

## 7.2 A formalization of failure

When a pair of expressions cannot be unified, MW's unification algorithm returns the special symbol 'nil.' My proof uses a simple polymorphic data type for failure. If $\alpha$ is a type, then the type $(\alpha)attempt$ includes the value FAILURE, and also SUCCESS $x$ for every $x : \alpha$. The LCF command to define *attempt* is simply

    *struct_axm* ("$:(\alpha)attempt$", 'strict',
        ['FAILURE', [];   'SUCCESS', ["$x : \alpha$"]]);;

A computation that sometimes returns a value of type $\alpha$ and sometimes fails can be formalized as a PPLAMBDA term of type $(\alpha)attempt$. The flow of control may depend on whether a computation fails or succeeds. We need to be able to test for the value FAILURE, then taking appropriate action, and to test for any value of the form SUCCESS $x$, then performing a calculation that may depend on $x$. This control structure is formalized as the function

$$\text{ATTEMPT\_THEN} : (\alpha)attempt \to (\beta \times (\alpha \to \beta)) \to \beta,$$

which satisfies

$$\text{ATTEMPT\_THEN FAILURE } (fail, succ) \equiv fail$$
$$\text{ATTEMPT\_THEN (SUCCESS } x) (fail, succ) \equiv succ\, x$$

Intuitively, evaluating ATTEMPT_THEN $z$ $(fail, succ)$ consists of evaluating $z$. If $z$ fails, then $fail$ is evaluated; if $z$ returns SUCCESS $x$, then the function invocation $succ\, x$ is evaluated. A typical use of ATTEMPT_THEN appears in the next section.

## 7.3 Unification

My definition of the unification algorithm looks different from MW's (§15), though it is essentially the same. Given a pair of terms, it considers all cases; either term may be a constant, variable, or combination. It is defined using an infix operator for unifying a variable with a term:

$$\text{ASSIGN} : var \to term \to ((var, term)alist)\ attempt$$

A variable may be unified with any term not containing it:

$v\ \text{ASSIGN}\ t \equiv ((\text{VAR}\, v)\ \text{OCCS}\, t) \Rightarrow \text{FAILURE} \mid \text{SUCCESS}(\text{ACONS}\ v\ t\ \text{ANIL})$

Unification is also written as an infix operator:

$$\text{UNIFY} : term \to term \to ((var, term)alist)\ attempt$$

A constant can be unified with itself, or with a variable:

$(\text{CONST}\, c)\ \text{UNIFY}\ (\text{CONST}\, c') \equiv c = c' \Rightarrow \text{SUCCESS}(\text{ANIL}) \mid \text{FAILURE}$
$(\text{CONST}\, c)\ \text{UNIFY}\ (\text{VAR}\, v) \equiv v\ \text{ASSIGN}\ (\text{CONST}\, c)$
$(\text{CONST}\, c)\ \text{UNIFY}\ (\text{COMB}\, t\, u) \equiv \text{FAILURE}$

Unification with a variable is already taken care of:

$$(\text{VAR}\, v)\ \text{UNIFY}\ t \equiv v\ \text{ASSIGN}\ t$$

Unifying a combination with a constant or variable is handled easily:

$(\text{COMB}\, t_1 t_2)\ \text{UNIFY}\ (\text{CONST}\, c) \equiv \text{FAILURE}$
$(\text{COMB}\, t_1 t_2)\ \text{UNIFY}\ (\text{VAR}\, v) \equiv v\ \text{ASSIGN}\ (\text{COMB}\, t_1 t_2)$

The hard case is the unification of the combination COMB $t_1 t_2$ with another, COMB $u_1 u_2$. First unify the left sons $t_1$ and $u_1$. If this fails, then return FAILURE;

otherwise apply the unifier, $s_1$, to the right sons $t_2$ and $u_2$. If the resulting terms cannot be unified then return FAILURE; otherwise use the unifier, $s_2$, to return the answer $s_1$ THENS $s_2$:

$$
\begin{aligned}
&(\text{COMB}\, t_1 t_2)\ \text{UNIFY}\ (\text{COMB}\, u_1 u_2) \equiv \\
&\quad \text{ATTEMPT\_THEN}\ (t_1\ \text{UNIFY}\ u_1) \\
&\qquad (\text{FAILURE}, \\
&\qquad\quad \lambda s_1.\ \text{ATTEMPT\_THEN}\ ((t_2\ \text{SUBST}\ s_1)\ \text{UNIFY}\ (u_2\ \text{SUBST}\ s_1)) \\
&\qquad\qquad (\text{FAILURE},\ \lambda s_2.\ \text{SUCCESS}(s_1\ \text{THENS}\ s_2)))
\end{aligned}
$$

## 7.4  Properties of substitutions and unifiers

Stating the correctness of unification requires an elaborate set of concepts. My formalization defines most of these as predicate symbols, with theories about the properties of each.

A substitution $s$ *unifies* the terms $t$ and $u$ if applying it makes them identical:

$$\text{UNIFIES}(s, t, u) \iff (t\ \text{SUBST}\ s) \equiv (u\ \text{SUBST}\ s)$$

The substitution $s_1$ is *more general* than $s_2$ whenever $s_2$ can be expressed as an instance of $s_1$ composed with some other substitution $r$:

$$\text{MORE\_GEN}(s_1, s_2) \iff \exists_D r.\ s_2 \equiv s_1\ \text{THENS}\ r$$

A unifier $s$ of terms $t$ and $u$ is *most general* whenever it is more general than any other unifier of $t$ and $u$:

$$
\begin{aligned}
&\text{MOST\_GEN\_UNIFIER}(s, t, u) \iff \\
&\quad \text{UNIFIES}(s, t, u) \land (\forall_D r.\ \text{UNIFIES}(r, t, u) \implies \text{MORE\_GEN}(s, r))
\end{aligned}
$$

Two terms *cannot be unified* if there exists no substitution that unifies them:

$$\text{CANT\_UNIFY}(t, u) \iff \forall_D s.\ \neg\, \text{UNIFIES}(s, t, u)$$

A substitution $s$ is *idempotent* whenever $s$ THENS $s \equiv s$. For the induction step of the verification to go through, the algorithm must produce only most-general, idempotent unifiers:

$$\text{BEST\_UNIFIER}(s, t, u) \iff (\text{MOST\_GEN\_UNIFIER}(s, t, u) \land s\ \text{THENS}\ s \equiv s)$$

Not every pair of terms $t$ and $u$ can be unified. A *best try* at unification is either a failure, if no unifier exists, or else is a success giving a "best unifier:"

$$
\begin{aligned}
&\text{BEST\_UNIFY\_TRY}(z, t, u) \iff \\
&\quad z \equiv \text{FAILURE}\ \land\ \text{CANT\_UNIFY}(t, u)\ \lor \\
&\quad \exists_D s.\ z \equiv \text{SUCCESS}\, s\ \land\ \text{BEST\_UNIFIER}(s, t, u)
\end{aligned}
$$

These predicates allow the concise formulation of theorems. The Most-General Unifier Corollary (MW §12) becomes

$$\text{MOST\_GEN\_UNIFIER}(s, t, u) \iff \forall_D r.(\text{UNIFIES}(r, t, u) \iff \text{MORE\_GEN}(s, r)).$$

The Variable Unifier Proposition, which MW (§12) prove in detail, is

(VAR $v$) OCCS $t \not\equiv$ TT $\implies$ MOST_GEN_UNIFIER(ACONS $v\ t$ ANIL, VAR $v$, $t$).

The statement of correctness is

BEST_UNIFY_TRY($t$ UNIFY $u$, $t$, $u$).

## 7.5 Special cases of the correctness of unification

After presenting the theory of substitutions, MW state the correctness of the unification algorithm as a single theorem whose proof covers sixteen pages. I have broken this into lemmas. For instance, it is impossible to unify two distinct constants, or unify a constant with a combination. The LCF proofs are trivial, using rewriting to unfold the definitions of the predicates:

$c_1 \not\equiv c_2 \implies$ CANT_UNIFY(CONST $c_1$, CONST $c_2$)

CANT_UNIFY(CONST $c$, COMB $t\ u$)

Proving that ASSIGN correctly tries to unify a variable $v$ with a term $t$ involves case analysis on whether $v$ occurs in $t$. Rewriting solves both cases:

BEST_UNIFY_TRY($v$ ASSIGN $t$, VAR $v$, $t$)

These results are enough to show that UNIFY is correct when one operand is a constant, by case analysis on the other operand:

BEST_UNIFY_TRY((CONST $c$) UNIFY $t$, CONST $c$, $t$)

MW take seven pages to discuss the unification of one combination with another (MW §20). The results can be summarized as three theorems about the unification of COMB $t_1 t_2$ with COMB $u_1 u_2$. Perhaps $t_1$ and $u_1$ cannot be unified:

CANT_UNIFY($t_1, u_1$) $\implies$ CANT_UNIFY(COMB $t_1 t_2$, COMB $u_1 u_2$)

Perhaps they can be unified, but $t_2$ and $u_2$ cannot be:

BEST_UNIFIER($s, t_1, u_1$) $\implies$
CANT_UNIFY($t_2$ SUBST $s$, $u_2$ SUBST $s$) $\implies$
CANT_UNIFY(COMB $t_1 t_2$, COMB $u_1 u_2$)

Perhaps both unifications succeed:

BEST_UNIFIER($s_1, t_1, u_1$) $\implies$
BEST_UNIFIER($s_2$, $t_2$ SUBST $s_1$, $u_2$ SUBST $s_1$) $\implies$
BEST_UNIFIER($s_1$ THENS $s_2$, COMB $t_1 t_2$, COMB $u_1 u_2$)

These last two proofs use rewriting to unfold definitions, alternating with simple tactics like DISCH_TAC for breaking up the goal. Both proofs use a tactical for grabbing an assumption, which is instantiated in a particular way, and moved into the goal to be rewritten. Despite occasional ugly steps, the proofs of all the theorems mentioned in this section consist of only about two dozen tactic invocations in total.

## 7.6 The well-founded induction

MW prove the correctness of unification by well-founded induction on pairs of expressions (MW §21). The well-founded ordering $\prec_{\text{un}}$ is expressed as a lexicographic combination involving the set of variables contained in a pair, and the occurs-in relation between expressions. The pair $(t_1, u_1)$ preceeds $(t_2, u_2)$ if the set of variables in $(t_1, u_1)$ is a proper subset of the set of variables in $(t_2, u_2)$, or if these sets are the same, and $t_1$ occurs in $t_2$. The unification algorithm terminates because, for every recursive call, it either reduces the set of variables contained in its arguments, or leaves this set the same while reducing the size of its arguments. I have formalized these ideas differently from MW.

The shrinking of the set of variables is expressed as a reduction in cardinality. The function CARDV denotes the number of distinct variables in a pair of terms. (Boyer and Moore would call CARDV a *measure function* [2].) It refers to several functions involving finite sets: VARS_OF, for the set of variables in a term, CARD, for cardinality, and the infix function UNION:

$$\text{CARDV}(t, u) \equiv \text{CARD}((\text{VARS\_OF}\, t)\; \text{UNION}\; (\text{VARS\_OF}\, u))$$

The occurs-in relation concerns whether one term occurs in another at any depth. The correctness proof requires this only at depth one, where $t$ occurs in COMB $t u$ and COMB $u t$. This allows formalizing the induction without using OCCS.

The Head Ordering Proposition justifies the first recursive call in UNIFY. Its formalization uses the infix function LT, which is the less-than relation for the natural numbers:

$$\text{CARDV}(t_1, u_1) \equiv \text{CARDV}(\text{COMB}\, t_1 t_2,\; \text{COMB}\, u_1 u_2)\, \vee$$
$$\text{CARDV}(t_1, u_1)\; \text{LT}\; \text{CARDV}(\text{COMB}\, t_1 t_2,\; \text{COMB}\, u_1 u_2) \equiv \text{TT}$$

If the second disjunct holds, then the set of variables in $(t_1, u_1)$ is smaller than the set of variables in (COMB $t_1 t_2$, COMB $u_1 u_2$). If the first disjunct holds, then the sets of variables have the same size, and $t_1$ occurs in COMB $t_1 t_2$. MW's version of the theorem, mixing my notation with theirs, is

$$(t_1, u_1) \prec_{\text{un}} (\text{COMB}\, t_1 t_2,\; \text{COMB}\, u_1 u_2)\,.$$

The Tail Ordering Proposition justifies the second recursive call, when the first has succeeded:

$$\text{BEST\_UNIFIER}(s, t_1, u_1) \Longrightarrow$$
$$(t_2\; \text{SUBST}\; s \equiv t_2\; \wedge$$
$$\text{CARDV}(t_2,\; u_2\; \text{SUBST}\; s) \equiv \text{CARDV}(\text{COMB}\, t_1 t_2,\; \text{COMB}\, u_1 u_2))\, \vee$$
$$\text{CARDV}(t_2\; \text{SUBST}\; s,\; u_2\; \text{SUBST}\; s)\; \text{LT}\; \text{CARDV}(\text{COMB}\, t_1 t_2,\; \text{COMB}\, u_1 u_2) \equiv \text{TT}$$

Either the idempotent unifier $s$ has no effect on the term $t_2$, or it eliminates a variable of $t_2$ or $u_2$. Incidentally, most of the arcane theory of idempotent substitutions is needed solely for proving the Tail Ordering Proposition. MW's formulation is much neater:

$$\text{BEST\_UNIFIER}(s, t_1, u_1) \Longrightarrow$$
$$(t_2\; \text{SUBST}\; s,\; u_2\; \text{SUBST}\; s) \prec_{\text{un}} (\text{COMB}\, t_1 t_2,\; \text{COMB}\, u_1 u_2)$$

Induction over certain well-founded relations can be reduced to structural induction [25]. Note how $\prec_{\text{un}}$ is built up from simpler relations. Induction over the lexicographic combination of two relations gives rise to nested inductions. Induction over the "immediate" occurs-in relation (of which OCCS is the transitive closure) is simply structural induction. The complication is that $\prec_{\text{un}}$ involves the less-than relation on the natural numbers, via the measure function CARDV. This requires structural induction over the natural numbers, and correctness must be stated as

$$\forall_D t\, u.\, \text{CARDV}(t, u)\ \text{LT}\ n \equiv \text{TT} \implies \text{BEST\_UNIFY\_TRY}(t\ \text{UNIFY}\ u,\ t,\ u)\ .$$

The first step is natural number induction on $n$. Rewriting solves the $\bot$ and 0 base cases. In the case for $n+1$, the induction hypothesis asserts that UNIFY is correct for any two terms that contain fewer than $n$ distinct variables. The next step is structural induction on the term $t$. Rewriting solves the $\bot$, constant, and variable base cases. In the combination case, $t$ has the form $\text{COMB}\, t_1 t_2$; two new induction hypotheses assert that UNIFY is correct when the left operand is $t_1$ or $t_2$.

In case analysis on $u$, the $\bot$, constant, and variable cases are again easy to prove. The only remaining case is where both $t$ and $u$ are combinations. Further case analyses consider whether unification succeeds in the first recursive call, then the second. The Head and Tail Ordering Propositions allow appeals to the three induction hypotheses, after yet more case analysis. Things are simpler for MW, thanks to $\prec_{\text{un}}$: their single induction hypothesis is easily instantiated using their forms of these Propositions.

The LCF proof is uncomfortably large: about two dozen tactic invocations, with complex combinations of tacticals, tactics, and inference rules. It appears feasible to implement an LCF package for well-founded induction, so that this proof would resemble MW's.

The antecedent involving CARDV is easy to remove. For any $t$ and $u$, the number $\text{CARDV}(t, u)$ is less than $\text{CARDV}(t, u) + 1$. Finally we get

$$\text{BEST\_UNIFY\_TRY}(t\ \text{UNIFY}\ u,\ t,\ u)\ .$$

# 8 Concluding comments

This LCF verification should certainly increase our confidence in the proof by Manna and Waldinger (MW). I found only a few errors, too trivial to list. We have seen only one complete proof, of the monotonicity of substitution. While its two-tactic LCF proof is surprisingly short, few proofs involve more than a dozen tactics; other proof checkers may require hundreds of commands to prove comparable theorems [11, 16]. Cohn has found that a six-tactic proof of parser correctness causes LCF to perform about eight hundred inferences [9]. Many LCF proofs are shorter than MW's detailed verbal proofs.

## 8.1 Logics of computation

Although parts of this proof expose weaknesses in LCF's automatic tools, the main cause of difficulty is termination. The logic PPLAMBDA is flexible enough to allow proving the termination of UNIFY, but at the cost of explicitly reasoning about $\bot$ everywhere. PPLAMBDA is most appropriate for proofs in denotational semantics [10, 20, 26]. Newer logics avoid $\bot$ and provide higher-order functions and quantifiers, but have no straightforward way of handling the unusual recursion of UNIFY [15, 18].

In a logic of total functions, there is a general approach to proving the termination of a function $f(x)$. Define the function $f'(x,n)$ to make the same recursive calls as $f(x)$, decrementing the bound $n$; if $n$ drops to zero, $f'$ returns the special token BTM. Then $f'(x,n)$ always terminates. To show that $f(x)$ terminates, find some $n$ such that $f'(x,n)$ differs from BTM. Boyer and Moore [3] use this approach to reason about partial functions like Lisp's EVAL. Though this concrete view of termination has a natural appeal, the bound $n$ appears to complicate proofs.

MW's logic has only total functions; they avoid the termination problem by *synthesizing* UNIFY instead of verifying it. They essentially prove

$$\forall_D t\, u.\ \exists_D z.\ \text{BEST\_UNIFY\_TRY}(z,t,u),$$

and then their methods produce a total function UNIFY. In Martin-Löf's Intuitionistic Type Theory [18], proving $\forall x.\, \exists y.\, P(x,y)$ always produces a computable, total function $f$ and a theorem $\forall x.\, P(x,f(x))$. This logic appears to be ideally suited for program synthesis [1].

## 8.2 Other work

Eriksson has synthesized a unification algorithm from a specification in the first-order predicate calculus [11]. The algorithm is expressed in Prolog style [8], as Horn clauses proved from axioms defining unification. The method guarantees partial correctness, since anything that can be proved from the Horn clauses can also be proved from the specification. Total correctness requires proving that the existence of output follows logically from the Horn clauses; these should be executed on a *complete* Horn clause interpreter, not the sort that loops given $P \Longrightarrow P$.

The derivation is simplified by not considering termination; also, Eriksson's algorithm does not report when two expressions cannot be unified. There is no need for well-founded induction or the theory of idempotent substitutions. Eriksson's theorems seem equivalent to those of section 7.5 that do not involve CANT_UNIFY. He has developed the proofs by hand in complete detail (a total of 2500 inference steps) and verified them by machine. This attractively simple scheme for program synthesis would obviously benefit from a more powerful theorem prover, such as LCF or Gordon's similar system, HOL [15].

## 8.3 Future developments

Cambridge LCF was developed as part of this verification. I began this proof using Edinburgh LCF, but soon felt a need for improvements, notably the logical symbols

∨ and ∃. The unification proof was an excellent benchmark for judging the merits of tactics and simplifiers: the most useful tools became part of Cambridge LCF.

Theories developed for one LCF proof have rarely been used for others; I do not know of any other use for idempotent substitutions. On the other hand, every LCF proof teaches us something about methods. The unification proof requires an understanding of many kinds of structural induction [24]. Deriving its well-founded induction rule involves a new set of techniques that has become one of my research interests [25]. Sokołowski's proof of the soundness of Hoare rules uses powerful generalizations of goals and tactics [26]. The constant accumulation of techniques means that future proofs can be more ambitious than this one.

*Acknowledgements:* In such a large project, it is hard to remember everyone who helped. M.J.C. Gordon was always available for discussion and questions. G. P. Huet and G. Cousineau invested considerable effort in the implementation of Cambridge LCF. I had valuable conversations with R. Burstall, R. Milner, and R. Waldinger. W.F. Clocksin and I.S. Dhingra read drafts of this paper; a referee made detailed comments.